\documentclass[floatfix,10pt,superscriptaddress,twocolumn,showpacs,showkeys,aps,prb,final]{revtex4-1}

\usepackage{graphicx}% Include figure files
\usepackage{dcolumn}% Align table columns on decimal point
\usepackage{subfigure}
\usepackage{bm}
\usepackage{ifpdf,braket}
\usepackage{epstopdf}

\ifpdf
\usepackage[pdftex,
        colorlinks=true,
        pdftitle={Variational principle applied to two-band superconductors},
        pdfauthor={Mauro M. Doria},
        baseurl={http://www.if.ufrj.br/~mmd},
        pdftoolbar=true,
        pdfmenubar=false,
        pdfpagemode=UseOutlines,
        pdfstartview=FitH,
        bookmarksopen=false
        ]{hyperref}
\fi

\begin{document}

\title{Zero helicity states in the LaAlO$_3$/SrTiO$_3$ interface}

\author{Edinardo I. B. Rodrigues}%
\affiliation{Departamento de F\'{\i}sica, Universidade Federal da Para\'{\i}ba, 58051-970 João Pessoa, Para\'{\i}ba, Brazil}
\author{Alfredo A. Vargas-Paredes}%
\affiliation{Departamento de F\'{\i}sica dos S\'{o}lidos,
Universidade Federal do Rio de Janeiro, 21941-972 Rio de Janeiro, Brazil}
\author{Mauro M. Doria}%
\affiliation{Dipartimento di Fisica, Universit\`a di Camerino, I-62032 Camerino, Italy}
\affiliation{ Departamento de F\'{\i}sica dos S\'{o}lidos, Universidade Federal do Rio de Janeiro, 21941-972 Rio de Janeiro, Brazil}%
\email{mmd@if.ufrj.br}
\noaffiliation
\author{Marco Cariglia}%
\affiliation{Departamento de F\'{\i}sica , Universidade Federal de
Ouro Preto, 35400-000 Ouro Preto Minas Gerais, Brazil}

\date{\today}

\begin{abstract}
We propose a kinetically driven mechanism based on the breaking of the  spatial reflection symmetry to describe the magnetic moment and the torque observed by Lu-Li et al. (Ref.~\onlinecite{luli11}) for the LaAlO$_3$/SrTiO$_3$ system. We find that the itinerant electrons are in a zero helicity state and predict the existence of charge inhomogeneities that cross the interface at constant rate. There is mass and thickness anisotropies between the two sides of the interface.
\end{abstract}

%74.20.De 	Phenomenological theories (two-fluid, Ginzburg-Landau, etc.)
%74.25.Ha 	Magnetic properties including vortex structures and related phenomena
%75.70.Kw 	Domain structure (including magnetic bubbles and vortices)
%75.70.Tj 	Spin-orbit effects

\pacs{74.20.De,74.25.Ha,75.70.Kw,75.70.Tj }

\maketitle

\textbf{Introduction}. --
The discovery~\cite{ohtomo04,thiel06} of a high-mobility electron gas  at the interface between the bulk insulators non-magnetic  LaAlO$_3$ (LAO) and SrTiO$_3$ (STO) unveiled unexpected  properties~\cite{fete12}, such as the coexistence of magnetism and superconductivity. The Rashba spin-orbit interaction~\cite{gorkov01} plays a pivotal role~\cite{caviglia10} acting on both the superconducting and the normal-state transport properties~\cite{shalom10}. There are signals of an underlying ferromagnetic order~\cite{dikin11} that appears as magnetic dipoles, but with no net magnetization~\cite{bert11}. Torque magnetometry~\cite{luli11} shows the existence of a magnetic moment parallel to the interface for a nearly perpendicular applied magnetic field. To reconcile the observed perpendicular (axis 3) and  parallel magnetic moments, theoretical proposals of elaborate magnetic patterns have been made, based on  spirals~\cite{banerjee13} and skyrmions~\cite{babaev14}, obtained from the ad hoc assumption of the Rashba term. A general two dimensional (2D) free energy has been suggested to describe the long-wavelength magnetism~\cite{li14} to find a lattice of skyrmions similar to helimagnets.
In this letter we propose that the magnetic moment and the torque observed in the  LAO/STO system are kinetically driven, which means that they stem from the standard 3D kinetic energy of itinerant electrons with no need for extra terms. Nevertheless the presence of an interface breaks the reflection  symmetry and gives rise to the Rashba term, which is then found to be contained in the 3D kinetic energy. As a consequence of this breaking the electronic confinement to the interface is {\it quasi} two-dimensional, the itinerant electrons are in a zero helicity state and there currents crossing and entering the interface constantly. The existence of a lattice of vortices and skyrmions in the interface follows from this zero helicity condition (ZHC). To describe the magnetic moment and the torque data of Ref.~\onlinecite{luli11} there must exist anisotropy~\cite{joshua12}, between the LAO and STO sides of the interface. The mass of surface carriers has been detected since long ago in superconducting thin films and wires deposited on a SrTiO$_3$~\cite{parage98}.

%------------------- torque --------------------------
The torque in the high-T$_c$ superconductors is known to be kinetically driven and is well described by the phenomenological London anisotropic theory~\cite{kogan88}. The London kinetic energy captures well the energetic unbalance caused by a supercurrent circulation that preferably remains along the easy mass plane, namely, the superconducting layers, for a tilted external field. As a result there is magnetic moment not oriented along the applied field and a consequent torque whose measurement can unveil new properties of the vortex state. The temperature dependence of the anisotropy has been used to give evidence of two band superconductivity in the layered compounds~\cite{bosma11}.
Torque measurements helped Lu-Li et al.~\cite{luli05} to claim that there is a rigid London order parameter above T$_c$ despite the loss of the Meissner effect in the high-T$_c$ compound $Bi_2Sr_2CaCu_2O_8$.
Theoretically it has been shown that the torque can show signals of vortex-vortex attraction, known to occur in the low tilted field regime of the high-T$_c$~\cite{doria94}. The unbalance caused by the coexistence of vortices of different lengths in granular superconductors leads to a torque~\cite{romaguera07}. In this letter we claim that the torque observed in the LAO/STO system is also kinetically driven. Fitting the magnetic moment and torque data of Ref.~\onlinecite{luli11} to the present theory gives information of the mass anisotropy and the effective thickness of the LAO and STO sides of the interface~\cite{joshua12}.
%-------------------FIGURAS-----------------------------
%----------------- figure magnetic moment - torque ----------------------
\begin{figure}
\includegraphics[width=0.8\linewidth]{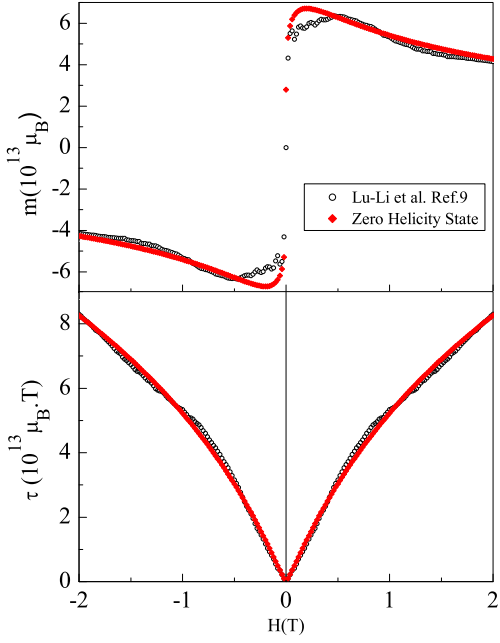}
\caption{(color online) Measured magnetic moment and torque data fitted by the present theory}\label{tmm}
\end{figure}
%-------------------------------------------------------------
%------------------------- campo magnetico local -------------------------
\begin{figure}[htp]
\centering
\subfigure[The local field $\delta h_3$ (perpendicular to the interface) \label{h3}]{\includegraphics[width=0.8\linewidth]{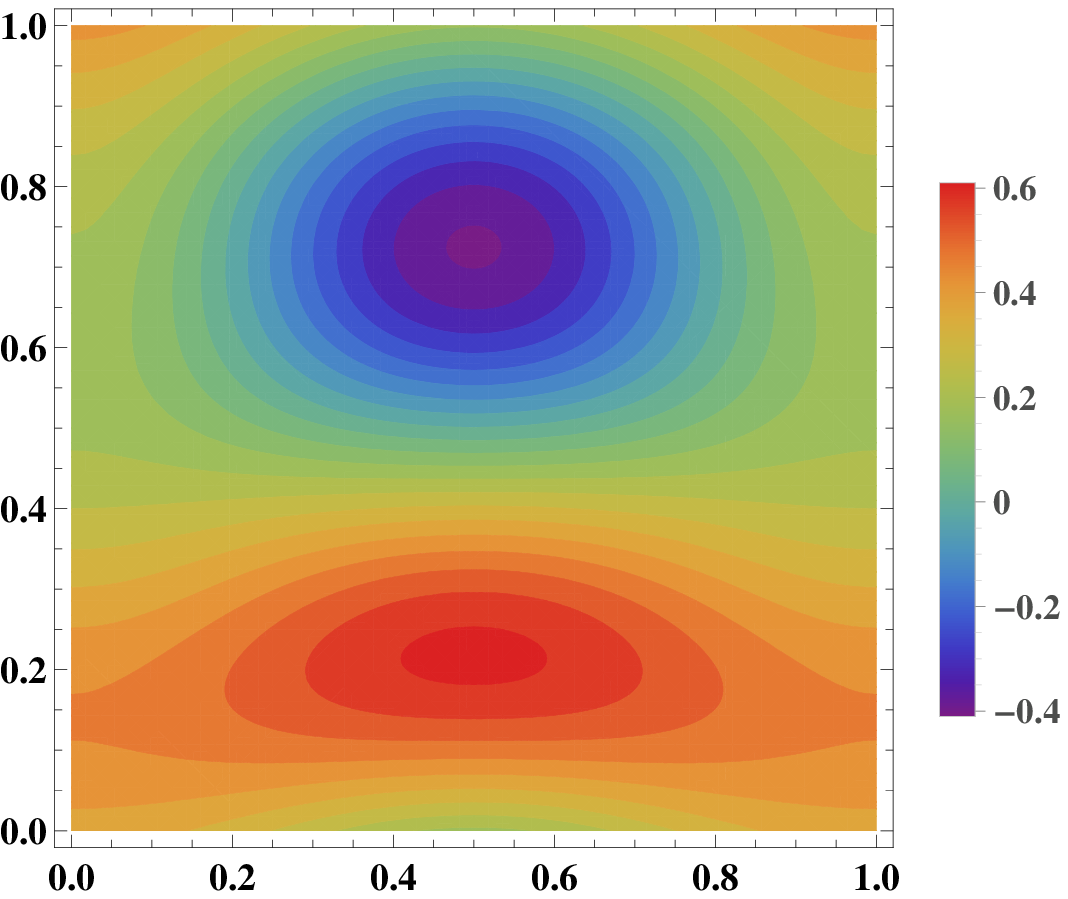}}\\
\subfigure[The local field $\delta \vec h_{\parallel}(0^+)$ (parallel and just above the interface). \label{h-above}]{\includegraphics[width=0.45\linewidth]{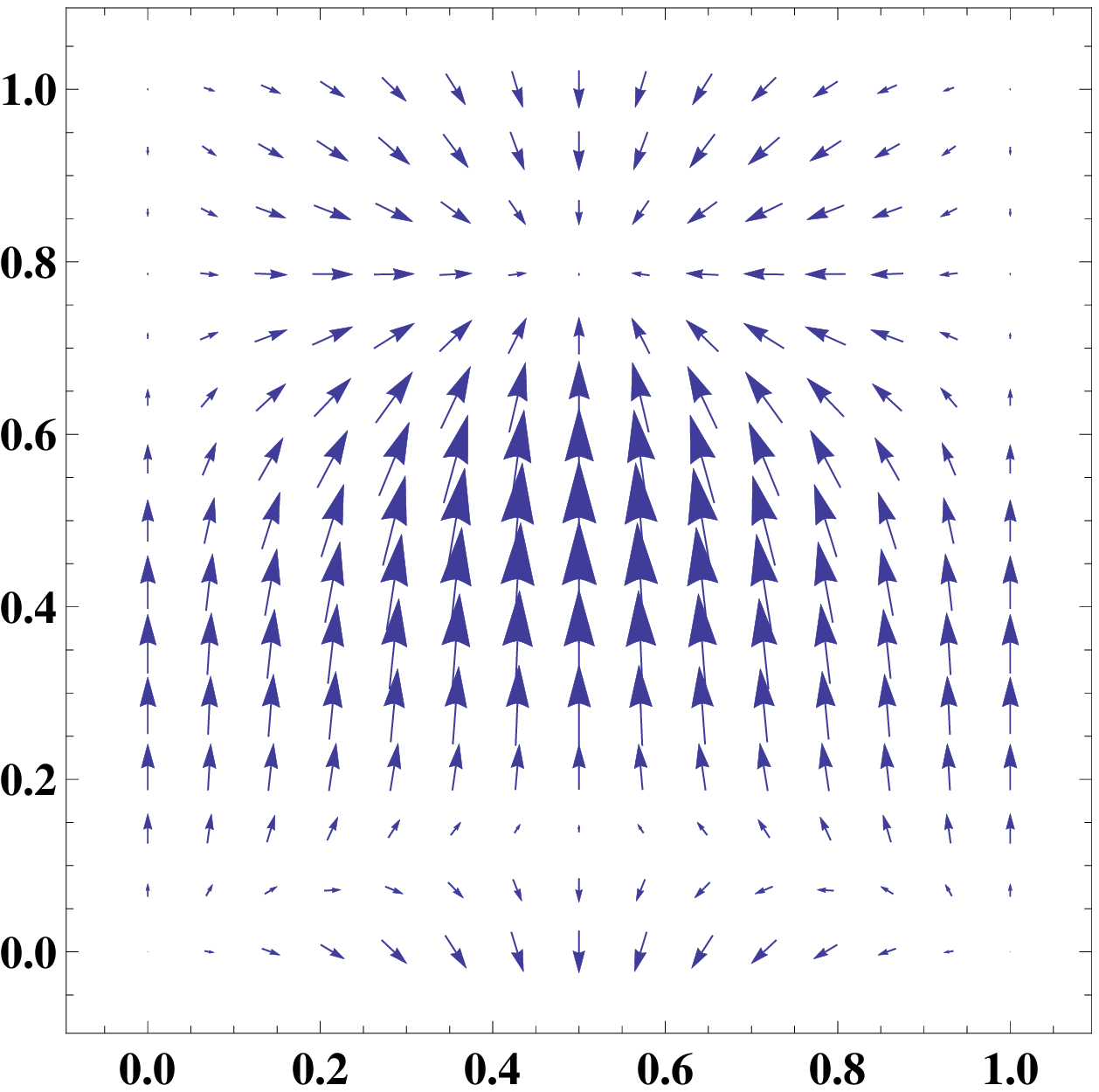}}\quad
\subfigure[The local field $\delta \vec h_{\parallel}(0^-)$ (parallel and just below the interface.)
\label{h-below}]{\includegraphics[width=0.45\linewidth]{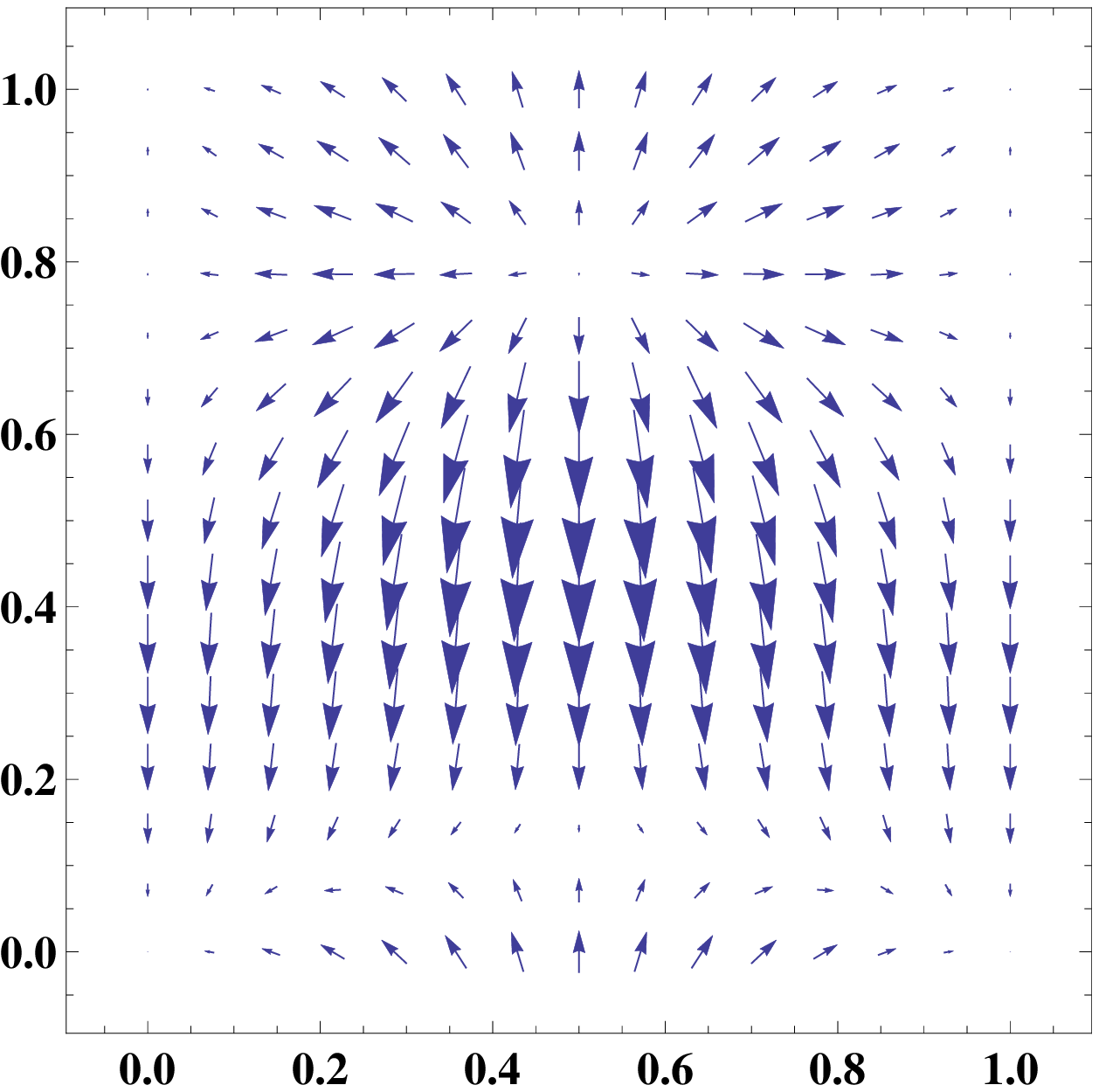}}
  \caption{The local magnetic field without the applied one is shown in the unit cell $(x_1/L,x_2/L)$,}
  \label{h-local}
\end{figure}
%--------------------------------------------------------------------
%----------- current, order parameter, charge density ---------------
\begin{figure*}[htp]
\centering
\subfigure[$-\vec \nabla_{\parallel} \cdot \vec J_s$ (the charge rate density)\label{cdw}]{\includegraphics[width=0.33\linewidth]{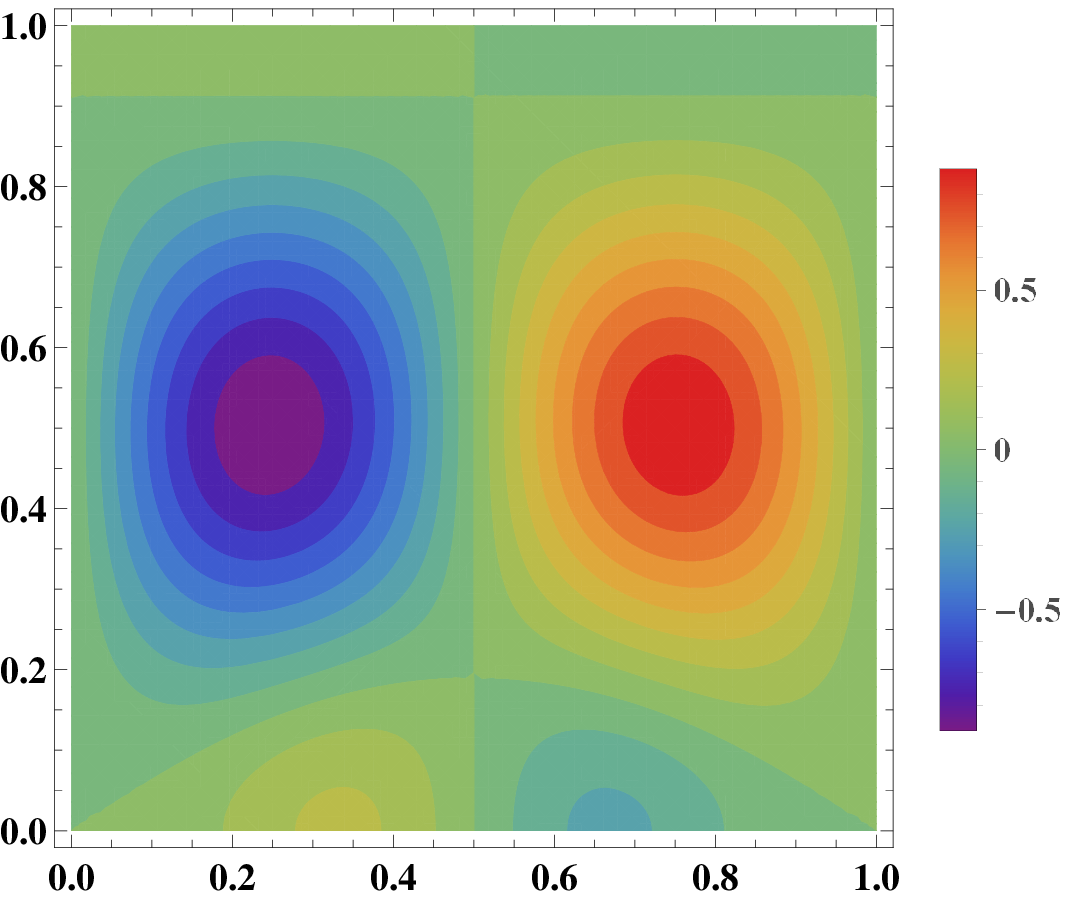}} \quad
\subfigure[$\vec J_s$ (the superficial current)\label{sc}]{\includegraphics[width=0.28\linewidth]{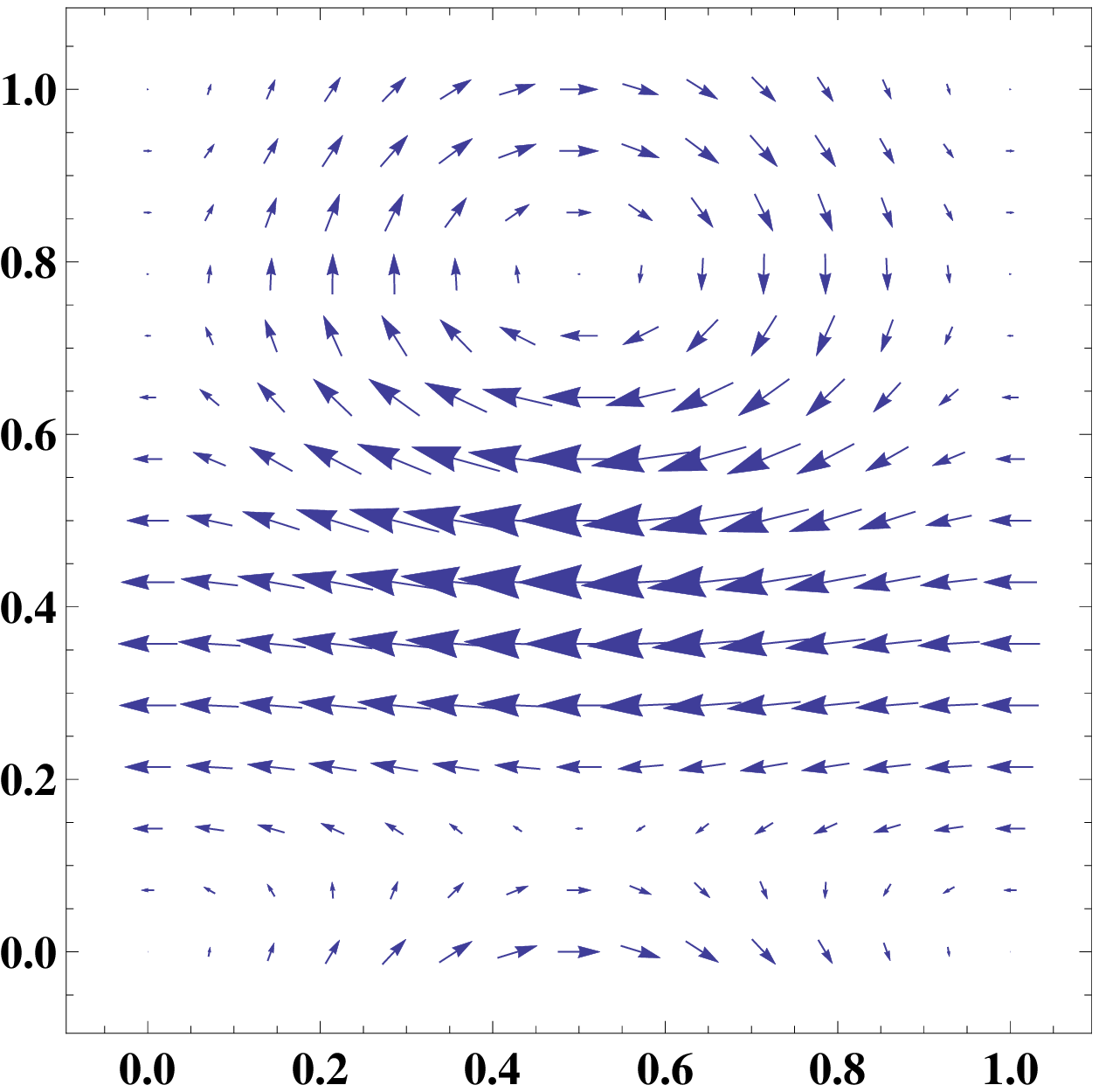}}\quad
\subfigure[$|\Psi|^2$ (the order parameter density )\label{op} ]{\includegraphics[width=0.33\linewidth]{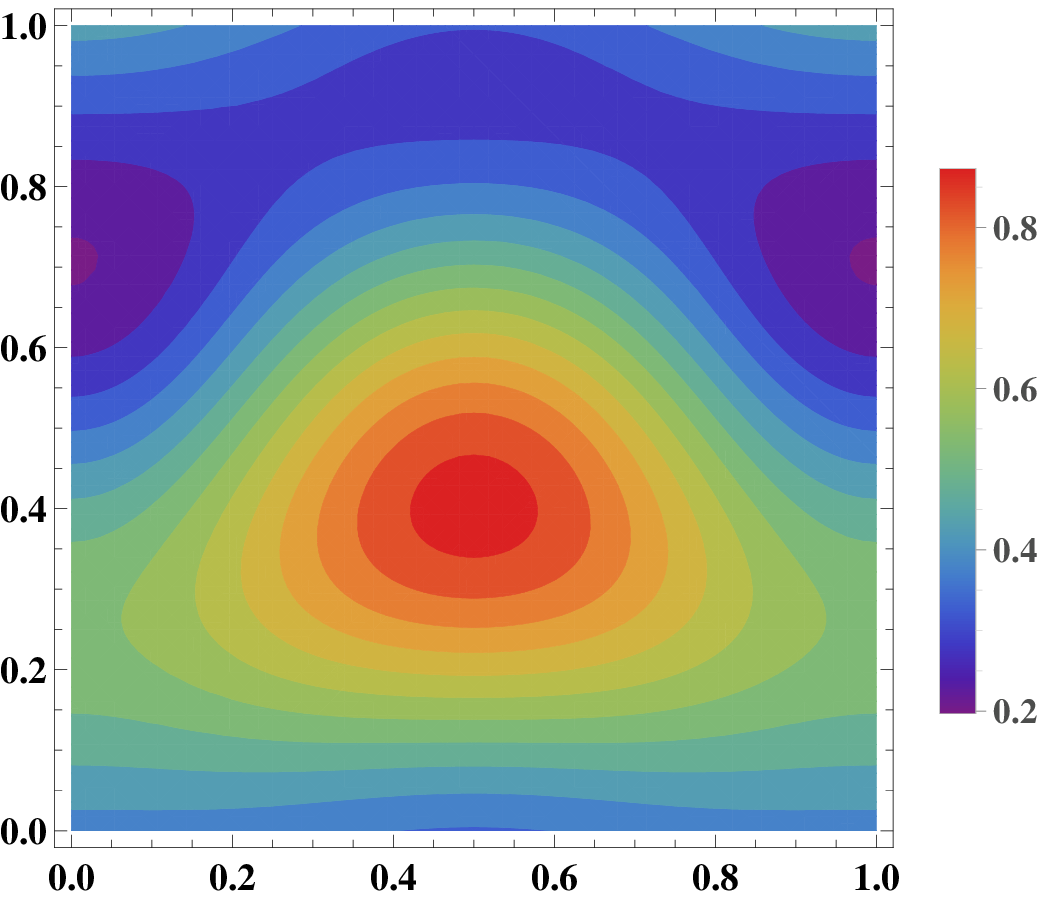}}
\caption{Properties at the interface $x_3=0$.}
  \label{sc-op-cdw}
\end{figure*}
%---------------------------------------------------------------------

\textbf{The theory}. --
Interestingly, according to Ref.~\onlinecite{luli11}, the magnetic ordering  and the superconducting state coexist and the torque remains nearly the same far above the superconducting critical temperature. We take this as an evidence of a kinetically driven torque, namely, that nor the condensate energy, below T$_c$, neither the interaction energy, above T$_c$, are relevant to the torque.
Therefore our starting point is the simplest possible order parameter theory composed by the sum of the 3D kinetic and field energies, $F=F_k+F_f$, where $F_f=\left \langle {\vec h}^2/8\pi \right \rangle$ takes the contribution of the local field $\vec h$ generated by the circulating currents. Notice that this starting point is exactly the same one taken to describe the torque of the high-T$_c$ superconductors~\cite{kogan88}. However a fundamental difference assumed here is the presence of two order parameters, each one associated to a distinct spin component.  The 3D kinetic energy is well known to be given by,
\begin{eqnarray}
F_k = \frac{1}{2m_0}\left \langle\left (\vec D'\Psi \right)^{\dag}\cdot \left (\vec D'\Psi \right) \right \rangle, \label{kin1}
\end{eqnarray}
where $\Psi$ is the two-component order parameter. The minimal coupling is given by $D'_k \equiv \sqrt{m_0/m_{(k)}} D_k$, $\vec D = (\hbar/i)\vec \nabla - (q/c)\vec A$, where $m_{(k)}$ are the mass parameters along the major axes, parallel to the interface ($k=1, 2$) and perpendicular to it ($k=3$). Uniaxial mass symmetry is assumed, such that the parallel masses are equal in both the LAO and STO sides, $m_{(1)}=m_{(2)}=m$, but the perpendicular masses are distinct in each side, and given by $m_{(3)}= M$ and $m_{(3)}=\bar  M$, respectively. The thickness of the LAO and STO sides are $d$ and $\bar d$, respectively, such that the active volume is $\delta V = A(d+\bar d)$, $A$ being the area of the interface. Therefore the integration volume comprises the LAO and STO volumes separately,
$\langle \cdots \rangle \equiv \langle \cdots \rangle_{x_3>0} + \langle \cdots \rangle_{x_3<0}$, where $\langle \cdots \rangle_{x_3>0}\equiv \int_{A} d^2 x_{\parallel} \int_0^d dx_3 \left (\cdots \right)/(A d)$ and $\langle \cdots \rangle_{x_3<0}\equiv \int_{A} d^2 x_{\parallel} \int_{-\bar d}^0 dx_3 \left ( \cdots \right)/(A \bar d)$.

Remarkably the 3D kinetic energy admits the following distinct but equivalent formulation~\cite{doria14}, suitable for the breaking of the spatial reversal symmetry,
\begin{eqnarray}\label{kin2}
&&F_k=
\frac{1}{2m_0} \left \langle\left\vert\vec{\sigma}\cdot\vec D'\Psi\right\vert^2+\frac{\hbar}{c}\vec{h} \cdot\Psi^\dag\vec\sigma'\Psi- \right . \nonumber
\\&&\left . -\frac{\hbar}{2}\vec \nabla'\left[\Psi^\dag\left(\vec \sigma \times
\vec D' \right)\Psi+c.c. \right] \right \rangle,
\end{eqnarray}
where $\nabla'_i \equiv \sqrt{m_0/m_{(i)}} \nabla_i$ and $\sigma'_i \equiv m_0/\sqrt{m_{(j)}m_{(k)}} \sigma_i$, $m_0$ being an arbitrary parameter. The Pauli matrices are $\sigma_i$  and $i\ne j \ne k$ take values among $1$, $2$, and $3$. Eq.(\ref{kin2}) splits the 3D kinetic energy into three contributions. The last term, proportional to $\vec \sigma \times
\vec D'$, is the Rashba term, only present at the edges, which here corresponds to just ($x_3=0^{+}$) above and ($x_3=0^{-}$) below the ($x_3=0$) interface. Along the interface we assume periodicity defined by a unit cell.
The helicity, $\vec{\sigma}\cdot\vec D'$, is a pseudo scalar, and consequently, the imposition that $\Psi$ makes it vanish triggers the  breaking of the spatial reversal symmetry.
Interestingly the ZHC yields the minimum electromagnetic energy since it exactly solves Amp\'ere's law, and fully determines the local magnetic field $\vec h = \vec \nabla \times \vec A$. To see this consider the electromagnetic current, known to be $\vec{J}=\left( q/m_0 \right)\left[\Psi^{\dag}\vec D'\Psi+ c.c. \right]$ in the traditional formulation of Eq.(\ref{kin1}). In our alternative formulation of  Eq.(\ref{kin2}), the current is $\vec{J}=\left ( q/2m_0 \right )\left \{\left[\Psi^{\dag}\vec{\sigma}\left(\vec{\sigma}.\vec D' \Psi\right)+c.c\right]-\hbar\vec \nabla' \times \Psi^{\dag}\vec{\sigma}\Psi \right \}$. Thus one obtains that,
\begin{eqnarray}
&&\vec{\sigma}\cdot\vec D'\, \Psi=0, \label{foe1} \mbox{and} \\
&&\delta \vec h \equiv \vec h - \vec H_{ext}= - \frac{h q}{m_0 c}\Psi^\dag  \vec \sigma' \, \Psi,
\label{foe2}
\end{eqnarray}
where $\delta \vec h$ is the local field without the external applied field.
A most important remark is that $\vec h$, obtained from Eq.(\ref{foe2}), is a solution of Maxwell´s equations, which means that $\vec \nabla \cdot \vec h = 0$. Thus $\vec h$ stream lines pierce the interface twice or none. The first case implies in  the existence of a superficial (2D) current density $\vec J_s$ at the interface since $\hat x_3 \times \big [\vec h(0^+)- \vec h(0^-)\big] = 4\pi\vec J_s(0)/c$. This current gives rise to the charge inhomogeneities along the interface since $\vec \nabla_{\parallel}\cdot \vec J_{s}\neq0$.
We obtain the magnetic moment and the torque under the two assumptions of a (i) small order parameter $\Psi \approx O(\epsilon)$ and of a (ii) small tilt angle $\theta$ between the applied field and the interface. The parameter $\epsilon$ controls the smallness of the fields, for instance,  from Eq.(\ref{foe2}) it follows that $\delta \vec h \approx O(\epsilon^2)$.
The perpendicular and parallel components of the applied field with respect to the interface, are given by $H_3=H_{ext}\cos\theta$ and $H_{\parallel}=H_{ext}\sin\theta$, respectively, such that $\vec H_{ext}=H_3\hat x_3+H_{\parallel}\hat e_{\parallel}$, where $\hat e_{\parallel}$ is a vector parallel to the interface.
We solve Eqs.(\ref{foe1}) and (\ref{foe2}) recursively to order $O(\epsilon^2)$, which means that $\Psi$ is obtained from  Eq.(\ref{foe1}) for $H_3\hat x_3$, with $H_{\parallel}$ discarded because of the assumption (ii). Next the inhomogeneous field $\delta \vec h$ is obtained from Eq.(\ref{foe2}) using the knwon $\Psi$. Further iterations of Eqs.(\ref{foe1}) and (\ref{foe2}) are not necessary to this lowest order $O(\epsilon^2)$.

The torque  directly follows from $\vec \tau = \vec m \times \vec H_{ext}$, once known the magnetic moment $\vec m =  V \vec M$, where
\begin{eqnarray} \label{mag}
\vec M = - \frac{\hbar q}{2m_0 c} \left \langle \Psi^\dag  \vec \sigma' \, \Psi \right \rangle.
\end{eqnarray}
is the magnetization. The magnetic induction is $\vec B =\langle \vec h \rangle$, and comparison of  Eq.(\ref{foe2}) with $\vec B=\vec H_{ext}+4\pi\vec M$ yields the above expression. Another way to obtain the torque is from the free energy, $\tau = -  V\partial F /\partial \theta$, and as we are restricted to the lowest order $O(\epsilon^2)$, only the following terms can contribute to it.
\begin{eqnarray}\label{kin3}
 F=  \frac{\hbar}{2 m_0 c} \vec{H}_{ext} \cdot \langle \Psi^{\dag}\vec\sigma'\Psi  \rangle +  \frac{\hbar^2}{4 m_0} \langle \nabla'^2 \left ( \Psi^{\dag}\Psi \right ) \rangle
\end{eqnarray}
Next we  show that the same torque can be obtained either from this free energy or the magnetic moment directly derived from Eqs.(\ref{foe1}) and (\ref{foe2}).
It can be shown that the last term of Eq.(\ref{kin3}) is proportional to $H_{3}$~\cite{edinardo15}, and so, it does not contribute to the torque under assumption (ii). A  way to see this proportionality is to consider that this last term vanishes for a spatially homogeneous state, namely $\Psi^{\dag}\Psi$ constant, and this is only achieved for $H_{3}=0$. Therefore the free energy becomes  $F \approx -\vec H_{ext} \cdot \vec M$ that can be reduced to $F \approx -\vec H_{\parallel} \cdot \vec M_{\parallel}$, where
$\vec M_{\parallel}= - (q\hbar/2m_0c)\langle \Psi^{\dag}\vec \sigma'_{\parallel}\Psi \rangle$, $\vec \sigma_{\parallel}'\equiv \sigma'_1 \hat x_1+\sigma'_1 \hat x_1$. There is no net magnetization perpendicular to the interface, $M_3=0$, in the present approach, which is in agreement with the experimental observations of Ref.~\onlinecite{bert11}. As shown below $M_{\parallel}$ only depends on $H_3$, and so the remain non-vanishing contribution under assumption (ii) comes from $\tau/V \approx - \partial \vec H_{\parallel}/\partial\theta \cdot \vec M_{\parallel}= H_3(q\hbar/2m_0c)\hat e_{\parallel}\cdot\langle \Psi^{\dag}\vec \sigma'_{\parallel}\Psi \rangle$, thus rendering the same torque of Eq.(\ref{mag}).

\textbf{The magnetic moment parallel to the interface}. -- The order parameter that satisfies the ZHC is obtained under the choice of gauge $A_3=0$ in Eq.(\ref{foe1}), and is,
\begin{equation}
\Psi = \sum_{n=1}^{\infty} \, c_n \,e^{-\sqrt{n}\frac{\vert x_3 \vert}{\scriptstyle{a}}}
\left( \begin{array}{c} \psi_{n}(x_1,x_2) \\ \frac{x_3}{\vert \scriptstyle{x_3} \vert}\psi_{n-1}(x_1,x_2) \end{array} \right)
\label{orderparam},
\end{equation}
It contains the $n^{th}$ Landau level functions, $\psi_n$, that appears twice in this series with the exception of $\psi_0$. They are normalized to $\int_{cell}(d^2x/L^2) \vert \psi_n\vert^2=1$, where the cell length is $L=\sqrt{\Phi_0/H_3}$.
The ZHC does not fully determines the order parameter  since any set of coefficients $c_n$ provide a new solution of Eq.(\ref{foe1}). We expect that residual interactions and higher order terms in the free energy determine the coefficients $c_n$ but we do not carry this procedure here. We take that the torque is not significantly sensitive to the free energy minimization, similarly to the case of the high-T$_c$ superconductor case. There the experimental torque cannot determine the optimal parameters of the vortex lattice.

The order parameter decays away from the interface and is distinct in each side of it. This decay is determined by $a=\eta(m/M)^{1/2}$ ($x_3>0$, LA0) and $a=\eta(m/\bar M)^{1/2}$, ($x_3<0$, ST0), respectively where $\eta= \sqrt{\Phi_0/4\pi H_{\perp}}$. Therefore the magnetic field controls both the periodicity at the interface, and also the evanescence perpendicular to it, through $L$ and $\eta$, respectively.
Indeed the exponential decay described by Eq.(\ref{orderparam}) sets a {\it quasi} 2D behavior away from the interface, that can be either 2D or 3D, and is controlled by $a$, and so, by $\eta$ and the anisotropy.
The magnetization is given by,
\begin{eqnarray}\label{magparallel}
\frac{\vec M_{\parallel}}{\mu_B}=-\sqrt{\frac{m}{M}}\left \langle \Psi^{\dag}\vec \sigma_{\parallel} \Psi \right \rangle_{x_3>0}- \sqrt{\frac{m}{\bar M}}\left \langle \Psi^{\dag}\vec \sigma_{\parallel} \Psi \right \rangle_{x_3<0}
\end{eqnarray}
where $\mu_B=\hbar q/2mc$, and we find that the two terms contribute oppositely,
\begin{eqnarray}
&& \left \langle \Psi^{\dag}\vec \sigma_{\parallel} \Psi \right \rangle_{x_3>0}=-i\hat x_{\parallel} \sum_{n=1}^{\infty} \, {c_n}^* c_{n+1}f_{(n)}\left ( r \right )-c.c.\\
&& \left \langle \Psi^{\dag}\vec \sigma_{\parallel} \Psi \right \rangle_{x_3<0}= i\hat x_{\parallel} \sum_{n=1}^{\infty} \, {c_n}^* c_{n+1}f_{(n)}\left ( \bar r \right )-c.c.
\end{eqnarray}
where $\hat x_{\parallel} = \left ( \hat x_1 + i\hat x_2 \right)$ and only the ratios $r \equiv (M /m)^{1/2}d /\eta$ and $\bar r \equiv(\bar M /m)^{1/2}\bar d/\eta$ matter for such averages and enter through the function $f_{(n)}\left ( z \right ) \equiv \left [ 1-\exp{\left(-z_n\right )}\right ]/z_n$, $z_n = \left( \sqrt{n+1}+\sqrt{n} \right)z$.
Notice the general property, $\vec M_{\parallel}\big[\big ( \frac{M}{m}\big)^{1/2},d,\big ( \frac{\bar M}{m}\big)^{1/2}, \bar d \big]=-\vec M_{\parallel}\big[ \big ( \frac{\bar M}{m}\big)^{1/2},\bar d,\big ( \frac{M}{m}\big)^{1/2}, d\big]$ which promptly shows that $\vec M_{\parallel}(-H_3)=-\vec M_{\parallel}(H_3)$ since $H_3 \rightarrow -H_3$ is equivalent to reverting the LAO and STO sides, namely, $d \leftrightarrow \bar d$ and $M \leftrightarrow \bar M$. Thus the present approach is in agreement with the reflection symmetry property  observed in Fig.(\ref{tmm}).
The asymptotic limits of $M_{\parallel} \sim \sqrt{H_{3}}$  and $M_{\parallel} \sim 1/\sqrt{H_{3}}$ are  a property of Eq.(\ref{magparallel}) for small and large fields, respectively and if added to the fact that $M_{\parallel}$ does not changes sign for $H_3>0$ , shows that the magnetization must reach a maximum, as seen in Fig.(\ref{tmm}).

%-------------------------------------------------------------
\textbf{Properties and comparison with the measured torque}. --  
The above general properties for the magnetic moment and torque are valid for any set of coefficients $c_n$ in Eq.(\ref{orderparam}) and allow for fitting of the data of Ref.~\onlinecite{luli11}. As pointed before $\psi_n$ appears twice, in consecutive doublets, therefore the simplest set able to capture the essence of this expansion corresponds to $c_1=c_2=\varepsilon$ real and $c_n=0$ for $n \le 3$. With respect to the choice of fitting parameters, the thickness of the LAO layer is  $\sim 5$ unit cells~\cite{luli11,fete12,annadi13} ($5a\approx 2.0 \, \mbox{}nm$, $a\approx 0.4 \, \mbox{nm}$).
Under these conditions we find for the best fitting the parameters $d = 1.8 \, \mbox{nm}$, $\bar d = 2.5 \, \mbox{nm}$, $(M/m)^{1/2} = 8.0$ and $(\bar M/m)^{1/2} = 8.5$ and is shown in Fig.(\ref{tmm}). From it we obtain that $\varepsilon \approx 9.4 \, \mbox{nm}^{-3/2}$. From fitting it follows that 
$V\varepsilon^2 = 1.6 \, 10^{15}$ and $V \approx 1.8 \, 10^{13} \mbox{nm}^3$. The volume is obtained by assuming  $0.3-0.4 \, \mu_B$ moments per LAO/STO cell area $a^2$, according to Ref.~\onlinecite{luli11}, and since there are approximately $10^{13} \, \mu_B$ moments in the system, total area of the interface is $A \approx 0.4 \, 10^{13} \mbox{nm}^2$, and the height is $d+\bar d = 4.3 \, \mbox{nm}$.
We notice that the maximum magnetic moment seen in Fig.(\ref{tmm}) corresponds to the parameters $r$ and $\bar r$ of order of a few units, thus showing that the exponential behavior of Eq.(\ref{orderparam}) sets the transition form  2D to 3D behavior.
From this theory it follows that a qualitative value for the local  field is $\delta h_{\parallel}  \approx 4\pi m_{\parallel}$ for a given $H_{ext}$.
For instance at the lowest applied field of $H_{ext} = 5\, \mbox{mT}$
the measured moment~\cite{luli11} is of $m_{\parallel} = 4.35 \, 10^{13} \mu_B$ and this gives that $\delta h_{\parallel}  \approx 70 \, \mbox{mT}$.

Figs.(\ref{h-local}) and (\ref{sc-op-cdw}) show properties of the state at $H_{ext}=0.5 \, \mbox{T}$.
The local magnetic field near to the interface is shown in Fig.(\ref{h-local}). Fig.(\ref{h3}) reveals positive and negative puddles of $\delta h_3$ at the interface, normalized to arbitrary units. This shows the existence of closed $\delta \vec h$ streamlines encircling the interface, confirmed by Figs. (\ref{h-above}) and (\ref{h-below}) which show $\delta \vec h_{\parallel}$ immediately above and below the interface, respecitively. The point $(0.5,0.8)$ is the skyrmion core as  $\vec \delta h$ stream lines cross from one side to the other of the interface to return elsewhere in the unit cell. There is  topologically stability for the skyrmion as the number,
$Q= (1/4\pi)\int_{x_3=0^+} \big [(\partial \hat
h/\partial x_1) \times (\partial \hat h/\partial x_2)
\big]\cdot \hat h \; d^2x$, is an integer ($Q=-2$). There is also vorticity carried in the phases of $\Psi$.
The 3D and 2D currents, $\vec J$ and $\vec J_s$, respectively, render a total divergenceless current at the interface too, which means that $\vec \nabla_{\parallel} \cdot \vec J_s + J_3(0^+)-J_3(0^-)=0$ and is interpreted as  $\vec \nabla_{\parallel} \cdot \vec J_s +\partial \sigma /\partial t=0$. Charge crosses the interface at constant rate $\partial \sigma /\partial t$ defined by the charge conservation forming positive and negative puddles as  seen in Fig.(\ref{cdw}). We obtain from the present model that  qualitatively $\partial \sigma /\partial t \sim \delta h_{\parallel} c/L$. Under the assumption that a charge unit $e$ crosses a nm$^2$ area this expression defines a rate that falls in the $THz$ range.
Fig.(\ref{sc}) depicts $\vec J_s$ and shows a circulation around the skyrmion center at the point $(0.5,0.8)$.  However this does not corresponds to a zero of the order parameter density, as shown in Fig.(\ref{op}). This is because the order parameter has two components and only one of them vanishes at this point meaning that indeed there is a vortex there, but associated to only one of the two available phases. In conclusion we have shown here that the ZHC explains the magnetic moment and torque data of Ref.~\onlinecite{luli11} from a kinetically driven mechanism.

\textbf{Acknowledgments}: We thank Prof. Dr. Lu Li
for kindly making available the data of Ref.~\onlinecite{luli11}.
M.M.D. acknowledges CNPq support from funding 23079.014992/2015-39, M. C. acknowledges FAPEMIG support from funding CEX-APQ-02164-14 and A. V-P. acknowledges CNPq support from funding 312535/2015-5 and FACEPE from DCR-APQ-0050-1.05/14.

\bibliography{reference}

\end{document}